\documentclass[%
reprint,
superscriptaddress,
amsmath,amssymb,
aip
]{revtex4-2}

\usepackage{graphicx}
\usepackage[dvipsnames]{xcolor}
\usepackage{dcolumn}
\usepackage{bm}
\usepackage[utf8]{inputenc}
\usepackage[T1]{fontenc}
\usepackage{mathptmx}

\draft 

\begin{document}

\title[Dong et al.]{Electronic structure of InSb (001), (110), and (111)B surfaces} 



\author{Jason T. Dong}
\affiliation{Materials Department, University of California, Santa Barbara, CA 93106}

\author{Hadass S. Inbar}
\affiliation{Materials Department, University of California, Santa Barbara, CA 93106}

\author{Mihir Pendharkar}
\altaffiliation{Present Address: Department of Materials Science and Engineering, Stanford University, Stanford, CA, 94305}
\affiliation{Department of Electrical and Computer Engineering, University of California, Santa Barbara, CA 93106}

\author{Teun A. J. van Schijndel}
\affiliation{Department of Electrical and Computer Engineering, University of California, Santa Barbara, CA 93106}

\author{Elliot C. Young}
\affiliation{Materials Department, University of California, Santa Barbara, CA 93106}

\author{Connor P. Dempsey}
\affiliation{Department of Electrical and Computer Engineering, University of California, Santa Barbara, CA 93106}

\author{Christopher J. Palmstr\o m}
\email[]{cjpalm@ucsb.edu}
\affiliation{Materials Department, University of California, Santa Barbara, CA 93106}
\affiliation{Department of Electrical and Computer Engineering, University of California, Santa Barbara, CA 93106}


\date{\today}

\begin{abstract}
The electronic structure of various (001), (110), and (111)B surfaces of n-type InSb were studied with scanning tunneling microscopy and spectroscopy. The InSb(111)B (3x1) surface reconstruction is determined to be a disordered (111)B (3x3) surface reconstruction. The surface Fermi-level of the In rich and the equal In:Sb (001), (110), and (111)B surface reconstructions was observed to be pinned near the valence band edge. This observed pinning is consistent with a charge neutrality level lying near the valence band maximum. Sb termination was observed to shift the surface Fermi-level position by up to $254 \pm 35$ meV towards the conduction band on the InSb (001) surface and $60 \pm 35$ meV towards the conduction band on the InSb(111)B surface. The surface Sb on the (001) can shift the surface from electron depletion to electron accumulation. We propose the shift in the Fermi-level pinning is due to charge transfer from Sb clusters on the Sb terminated surfaces. Additionally, many sub-gap states were observed for the (111)B (3x1) surface, which are attributed to the disordered nature of this surface. This work demonstrates the tuning of the Fermi-level pinning position of InSb surfaces with Sb termination.

\end{abstract}

\pacs{}

\maketitle 

\section{Introduction}

The breaking of translational symmetry at a surface or interface of a crystal allows evanescent waves to exist within the gap of a semiconductor, generating surface states \cite{Heine1965}. Semiconductor surface states have a strong influence on the properties of semiconductor devices. At sufficiently high surface state densities, these surface states can pin the Fermi-level of a semiconductor surface regardless of the semiconductor's bulk Fermi-level position and inhibit the control of the Schottky barrier height of the metal-semiconductor interface \cite{Bardeen1947}. Additionally, semiconductor surface states can serve as a source of surface recombination, which can negatively impact device performance \cite{Garrett1955}.

Scanning tunneling microscopy (STM) combined with scanning tunneling spectroscopy (STS) is a powerful tool to study the electronic properties of semiconductor surfaces \cite{Feenstra1987,Feenstra1988}. From a differential conductance (dI/dV) spectra of a semiconductor surface, semiconductor surface states can be directly measured. Additionally, if the conduction and valance band edges of the semiconductor surface are clearly distinguishable from any surface states, the Fermi-level position with respect to the band edges of the surface can be determined once tip induced band bending effects are accounted for \cite{Feenstra2007}. The aforementioned measurements can be performed with high spatial and energy resolution, making STM/S a useful tool to investigate the properties of semiconductor surfaces.

InSb is a III-V zincblende semiconductor with a small ($\sim$$0.2$ eV) band gap and a high ($\sim$$80,000$ cm\textsuperscript{2}V\textsuperscript{-1}s\textsuperscript{-1}) room temperature electron mobility \cite{Litwin-Staszewska1981}. The small band gap of InSb has enabled the use of InSb for infrared detection\cite{Shkedy2011,Ueno2013}. InSb has been explored as a material for high-speed and low power transistors due to its high electron mobility\cite{Datta2005,Radosavljevic2008,Alamo2011}. Recently, InSb has attracted interest as a candidate material for terahertz electronics\cite{Chochol2016,Ma2016}, and as a building block for topological quantum computers\cite{Lutchyn2018,Pendharkar2021}. In the application of topological quantum compuing, the Fermi-level pinning of InSb is an especially important parameter. In the proposals for implementing InSb in topological quantum computing, InSb is interfaced with superconductors to form hybrid semiconductor/superconductor heterostructures. The Fermi-level pinning is predicted to influence the degree of hybridization between the two materials and thus the properties of these heterostructures \cite{Mikkelsen2018}. To better engineer InSb based devices, understanding the surface electronic properties of InSb is essential.

While many of the clean InSb surfaces have been imaged with STM\cite{Jones1994a,McConville1994,Goryl2007,Davis1999,Jones1994,Varekamp1996a,Schweitzer1993,Whitman1990,Eguchi2002,Wever1994,Bjorkqvist1996,Nishizawa1997,Makela2018}, there have been only a few studies on the surface electronic structure of InSb with STS\cite{Whitman1990,Feenstra1991,Makela2018}. Most STS studies of InSb have been on the cleaved (110) (1x1) surface\cite{Whitman1990,Feenstra1991}, which have demonstrated the cleaved surface can be unpinned and devoid of surface states within the semiconductor band gap. In this work, the electronic structure of InSb (001), (110), and (111)B wafers prepared by molecular beam epitaxy and atomic hydrogen cleaned is studied by low temperature STM/S. The findings yield further insights into the nature of InSb surfaces.

\section{Methods}

\subsection{Sample Preparation}
Unintentionally doped InSb (001), (110), and (111)B wafers (Wafer Technology Ltd.) were used as the substrates for this study. The InSb wafers had an unintentional n-type doping of $\sim1\times10^{14}$ cm\textsuperscript{-3}. After loading into the vacuum chamber, the native oxide of the InSb wafers was removed with \textit{in-situ} atomic hydrogen cleaning. To hydrogen clean the samples, the samples were heated to 380\textsuperscript{o}C, as measured by a thermocouple near the sample, and exposed to a chamber pressure of $5 \times 10^{-6}$ Torr of atomic hydrogen, as measured by an ion gauge on the chamber. The atomic hydrogen was generated by flowing hydrogen gas through a leak valve and into a thermal cracker (Dr. Eberl MBE-Komponenten GmbH), which was operated at a temperature of 1700\textsuperscript{o}C. This hydrogen cleaning step resulted in the c(8x2), (1x1), and (3x3) surface reconstructions for the (001), (110), and (111)B surfaces, respectively. To obtain different surface reconstructions, the samples were transferred \textit{in-situ} into a molecular beam epitaxy chamber with a base pressure $< 1 \times 10^{-10}$ Torr, where a 20 - 100 nm thick unintentionally doped InSb layer was grown. The final surface reconstruction was determined by annealing the surface at different Sb overpressures and monitoring the surface reconstruction with reflection high energy electron diffraction. After surface preparation, the samples were rapidly cooled and then transferred \textit{in-situ} into a STM.

\subsection{Scanning Tunneling Microscopy and Spectroscopy}

\textit{In-situ} scanning tunneling microscopy was performed with an Omicron Low Temperatuer STM at 4.6 K or 77 K with a base pressure $< 4 \times 10^{-11}$ Torr. Electrochemically etched tungsten tips or mechanically cut PtIr tips were used for the STM experiments. The native oxide on the tungsten tips was removed with electron beam heating. All tips were characterized on gold prior to the STM experiments. Scanning tunneling spectroscopy was performed at constant tip height. The dI/dV was measured with a lock-in amplifier, with an AC modulation frequency of  $\sim$ 1 kHz and modulation voltage of 5-10 mV.

\section{Results and Discussion}

\begin{figure}[t]
\includegraphics[width=0.5\textwidth]{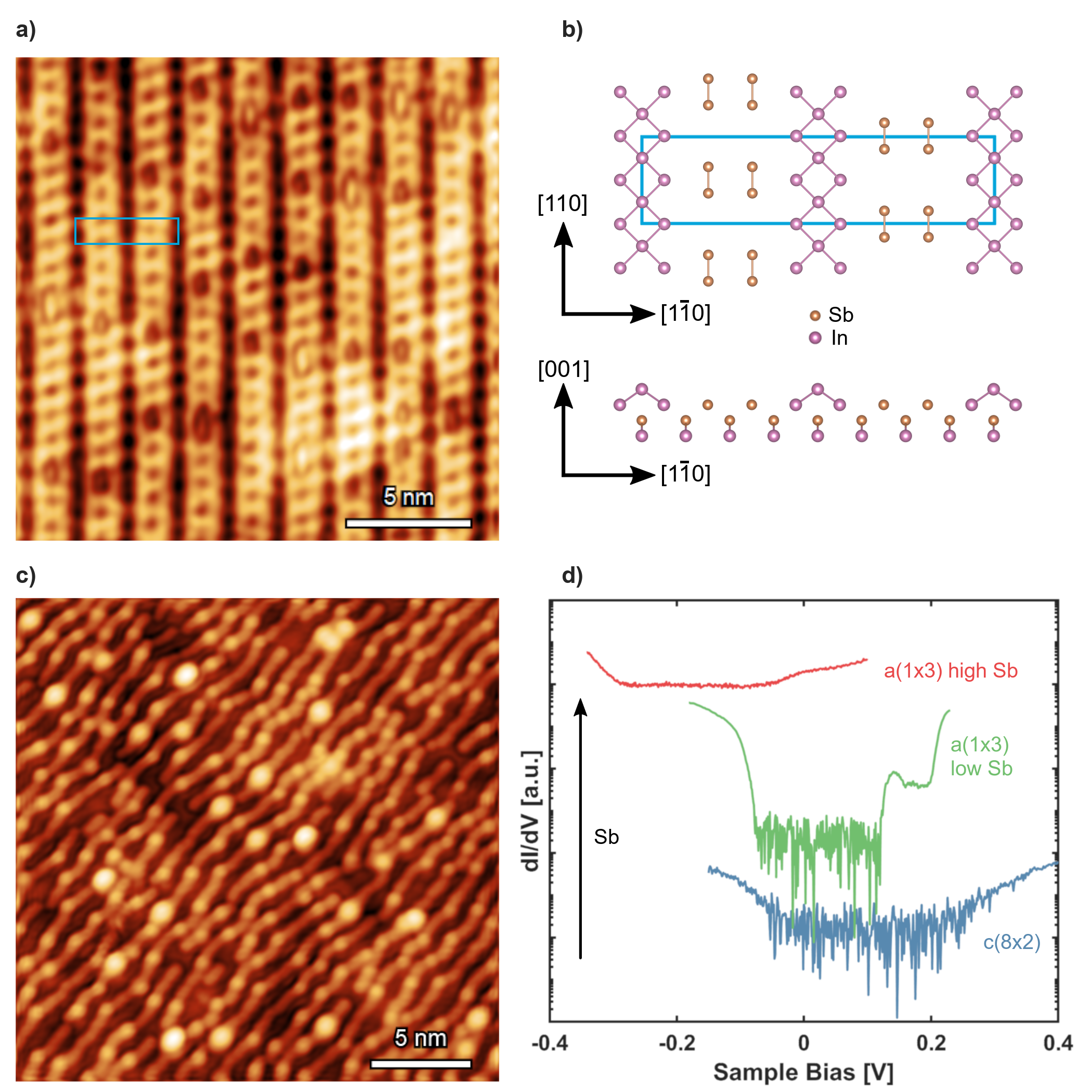}
\caption{\label{fig:1}STM/S of the InSb (001) surface. (a) Filled state STM image of the In rich c(8x2) surface reconstruction (V = -0.7 V, I = 1 nA). Surface reconstruction unit cell is outlined with a blue box. The bright spots in the image are Sb dimers, and the round defects are domain boundaries in the surface reconstruction. (b) top and side view of the atomic model of the c(8x2) surface reconstruction\cite{Davis1999}. The surface reconstruction unit cell of the atomic model is outlined in blue. (c) Empty state STM image of the Sb rich a(1x3) surface reconstruction (V = 1.0 V, I = 200 pA). The lines are chains of Sb atoms and the larger dots on the sample are likely excess Sb.  (d) STS spectra of In rich c(8x2) surface, a(1x3) surface with low Sb, and a(1x3) surface with high Sb. Spectra plotted on a semi-log scale and vertically offset.}
\end{figure}

\subsection{(001) Surface}

A filled states STM topographic image of the (001) centered (8x2) surface is shown in Fig. \ref{fig:1}(a). This centered (8x2) surface, or c(8x2) surface reconstruction, is an In rich surface of the InSb (001) surface \cite{Liu1994}. Visible in the STM image are Sb dimers. Columns composed of two Sb dimers are present. The c(8x2) surface is composed of pairs of Sb dimers separated by rows of In atoms. The surface reconstruction unit cell is outlined in blue. A ball and stick model\cite{Davis1999} of the c(8x2) surface reconstruction is shown in Fig. \ref{fig:1}(b) with the unit cell outlined in blue. The STM images of the c(8x2) surface are consistent with prior studies of the c(8x2) surface \cite{McConville1994,Goryl2007,Davis1999,Jones1994,Varekamp1996a,Schweitzer1993}. Additionally, round defects are present in the STM image. These defects are domain boundaries in the surface reconstruction. The domain boundaries are formed when different domains of Sb dimer stacking intersect. These domain boundaries are observed at low temperatures and are discussed at length elsewhere \cite{Goryl2007}.

Fig. \ref{fig:1}(c) shows a empty states STM topographic image of the (001) asymmetric (1x3) surface, or a(1x3) surface. The a(1x3) is a Sb-rich InSb(001) surface reconstruction\cite{Liu1994}. The surface has a "wormlike" morphology with bright dots on top of the "worms". The "wormlike" chains are disordered rows of Sb atoms. The bright dots on top of the chains are likely excess Sb clusters on the surface. This observed surface morphology is similar to other reports on the InSb a(1x3) surface\cite{McConville1994,Jones1994}. It was proposed that the a(1x3) surface reconstruction is (2x4) surface reconstruction composed of different polymorphs of the (2x4) surface reconstruction \cite{Liu1994}. However, an atomic model of the a(1x3) surface has yet to be directly confirmed by STM. 

The dI/dV spectra, or STS spectra, of the InSb (001) c(8x2) and a(1x3) surfaces is shown in Fig. \ref{fig:1}(d). Two a(1x3) surface where prepared with low and high Sb. The low Sb a(1x3) surface was prepared by supplying enough Sb to change the surface reconstruction from the c(8x2) to the a(1x3). The high Sb a(1x3) surface was prepared by supplying additional Sb beyond what was supplied to obtain the a(1x3) surface. The spectra are vertically offset for clarity. The spectra of the c(8x2) surface reconstruction is shown in blue. This spectra is devoid of subgap states and the valence band edge is $36 \pm 25$ meV below the Fermi-level (0V). The conduction band edge is $213 \pm 25$ meV above the Fermi-level. The band gap energy is determined by subtracting the valence band edge energy from the conduction band edge energy. The measured band gap is $249 \pm 35$ meV for the c(8x2) surface, which is in good agreement with the expected low temperature band gap of 240 meV\cite{Littler1985}. The good agreement between the measured band gaps and the expected band gaps is indicative of no significant tip induced band bending, and thus the voltage in the STS spectra can be directly interpreted as the energy, eV. All STS spectra presented in this report have measured band gaps of InSb that are in good agreement with the low temperature band gap of InSb. The spectra of the low Sb a(1x3) surface is shown in green, the valence band edge is $74 \pm 25$ meV below the Fermi-level and the conduction band edge is $197 \pm 25$ meV above the Fermi-level. A subgap state at $126 \pm 25$ meV is present in the a(1x3) low Sb spectra, this state is potentially from the excess Sb on the surface. Surface Sb islands are reported to posses states within the band gap of III-V semiconductors\cite{Feenstra1988}. The spectra of the high Sb a(1x3) surface is shown in red, the valance band edge is $290 \pm 25$ meV below the Fermi-level and the conduction band edge is $63 \pm 25$ meV below the Fermi-level, indicating the surface is in electron accumulation. There are no surface states within the band gap of the high Sb-rich a(1x3) surface. Through adding additional Sb on the (001) surface of InSb the Fermi-level shifts towards the conduction band by $254 \pm 35$ meV, relative to the valence band edge. Hence, Sb can be used to tune the surface Fermi-level position from close to the valence band to the conduction band.

\begin{figure*}[t!]
\includegraphics[width=0.8\textwidth]{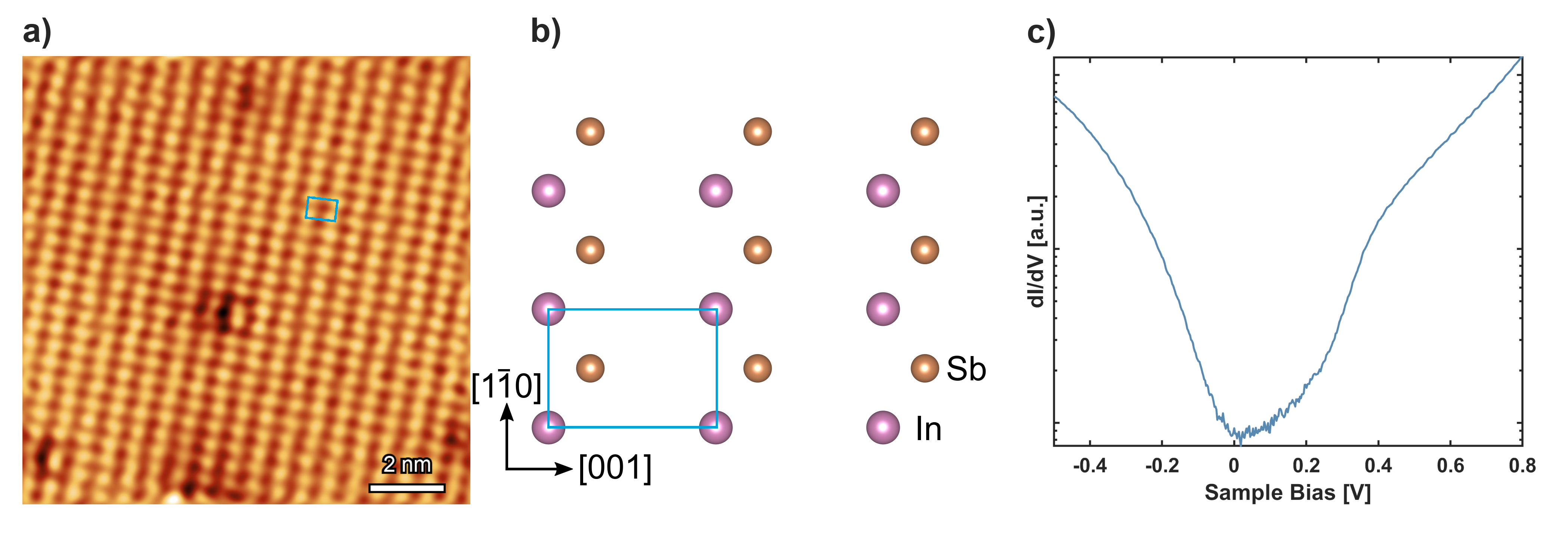}
\caption{\label{fig:2}STM/S of the InSb(110) (1x1) surface. (a) Atomic resolution empty state image of the (1x1) surface reconstructions (V = 1.5 V, I = 1 nA). The In atoms of the surface reconstruction are imaged. The surface reconstruction unit cell is outlined in blue. (b) Atomic model of the (1x1) surface reconstruction, the surface unit cell is outlined in blue. (c) STS spectra of the (1x1) surface reconstruction.}
\end{figure*}

\begin{figure*}[]
\includegraphics[width=0.8\textwidth]{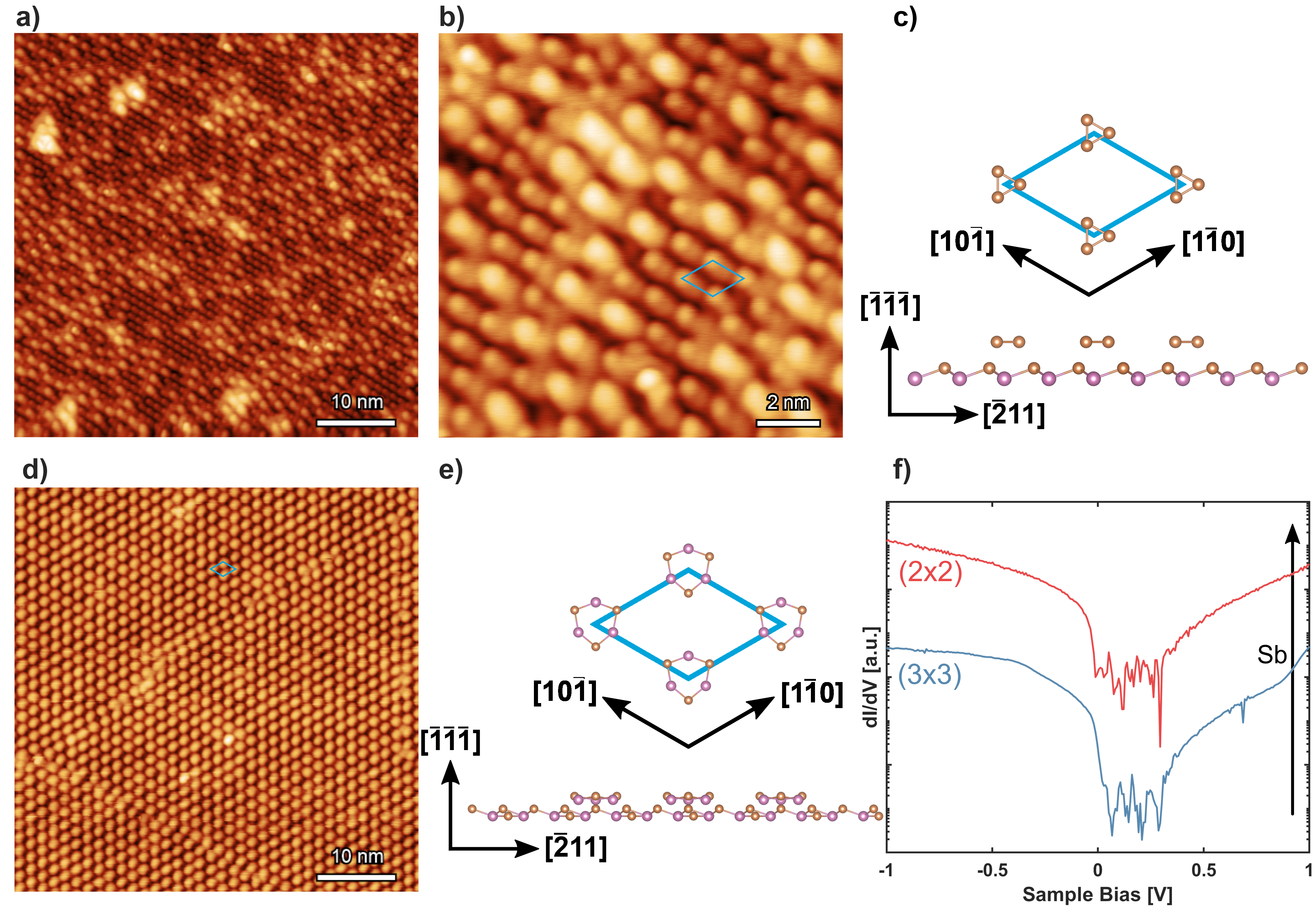}
\caption{\label{fig:3} (a) Empty state STM image of the InSb(111)B (2x2) surface reconstruction (V = 1.0 V, I = 70 pA). (b) Higher resolution STM image of the (2x2) surface reconstruction, with the surface reconstruction unit cell outlined in blue (V = 1.0 V, I = 70 pA). Sb trimers and larger Sb clusters are visible. (c) Atomic model of the (2x2) surface reconstruction\cite{Eguchi2002}, which is comprised of Sb trimers. The unit cell is outlined in blue. (d) Empty state STM image of the InSb(111)B (3x3) surface reconstruction (V = 1.0 V, I = 400 pA). Hexamers are visible in this STM image and the surface reconstruction unit cell is outlined in blue. (e) Atomic model of the (3x3) surface\cite{Wever1994}, surface is comprised of hexamers, which contain a ring of alternating In and Sb atoms. The unit cell is outlined in blue. (f) STS spectra of the (2x2) and (3x3) surface, spectra are vertically offset.}
\end{figure*}

\subsection{(110) Surface}

Fig. \ref{fig:2}(a) shows an empty state STM image of the (110) (1x1) surface. Visible in the image are the In atoms present on the surface. The In atoms form a rectangular lattice characteristic of the (1x1) surface reconstruction. The surface unit cell is outlined in blue. Fig. \ref{fig:2}(b) is the atomic model of the InSb (1x1)\cite{Whitman1990}, which is a surface comprised of a rectangular lattice of Sb and In atoms and natively has a 1:1 ratio of Sb:In. Point defects are also present in the STM image of the (1x1) surface. The observed (1x1) STM images are consistent with prior studies of the (1x1) surface \cite{Whitman1990}. The STS spectra of the (1x1) surface is shown in \ref{fig:2}(c). The valence band edge is $22 \pm 25$ meV below the Fermi-level and the conduction band edge $249 \pm 25$ meV above the Fermi-level. Other STS studies of cleaved InSb (110) (1x1) reveal an unpinned surface\cite{Feenstra1991,Whitman1990}. The observed pinning near the valence band maximum of the (1x1) surface is likely due to a larger number of step edges and point defects being present in the (110) (1x1) surface prepared by atomic hydrogen cleaning. Similar defect induced Fermi-level pinning of the (110) (1x1) surface has been observed before in GaAs\cite{Guichar1976}. There is a shoulder of states present in the band gap. These states are potentially dopant induced states, which have been studied extensively elsewhere\cite{Ishida2009}, or states due to the point defects on the surface.

\subsection{(111)B Surface}
An empty states STM image of the InSb (111)B (2x2) surface is shown in Fig. \ref{fig:3}(a). A higher resolution scan of the surface is shown in Fig. \ref{fig:3}(b). Visible in the STM image are the surface trimers and additional spots. The surface reconstruction unit cell is outlined in blue. The (111)B (2x2) surface is a Sb rich surface of InSb(111)B \cite{Noreika1981}. An atomic model of the (2x2) surface\cite{Eguchi2002} is shown in \ref{fig:3}(c). The surface is composed of Sb trimers which comprise form the main visible features in the STM image. The STM images are consistent with prior STM studies of the InSb(111)B (2x2) surface\cite{Eguchi2002}. The larger spots visible in the image are likely clusters of excesss Sb on the surface. Fig. \ref{fig:3}(d) shows an empty state STM image of the InSb (111)B (3x3) surface reconstruction. Hexamers are visible in the STM image. The surface reconstruction unit cell is outlined in blue. There are multiply polytypes of the (3x3) surface reconstruction hexamers, with varying In:Sb ratios and shapes\cite{Wever1994}. Based on the surface morphology, the particular hexamer polytype observed after hydrogen cleaning is determined to be a buckled ring of alternating Sb and In atoms. This surface nominally has an equal In:Sb ratio. The atomic model of the surface reconstruction\cite{Wever1994} is shown in Fig. \ref{fig:3}(e), with the unit cell outlined in blue. The STS spectra of the (3x3) and (2x2) surface reconstructions is shown in \ref{fig:3}(f). The spectra are vertically offset for visual clarity, with the red spectra corresponding to the (2x2) suface and the blue spectra corresponding to the (3x3) surface. Both spectra show clearly resolvable band edges and a bandgap devoid of subgap states. The valence band edge of the (2x2) $4 \pm 25$ meV above the Fermi-level, and the valence band edge of the (3x3) surface is $64 \pm 25$ meV above the Fermi-level. The conduction band edge of the (2x2) is $275 \pm 48$ meV above the Fermi-level, and (3x3) surface conduction band edge is $285 \pm 25$ meV above the Fermi-level. The (3x3) surface is in slight hole accumulation. There is a small $60 \pm 35$ meV upwards shift of the Fermi-level towards the conduction band upon Sb deposition, relative to the valence band edge. 

\begin{figure*}[]
\includegraphics[width=0.8\textwidth]{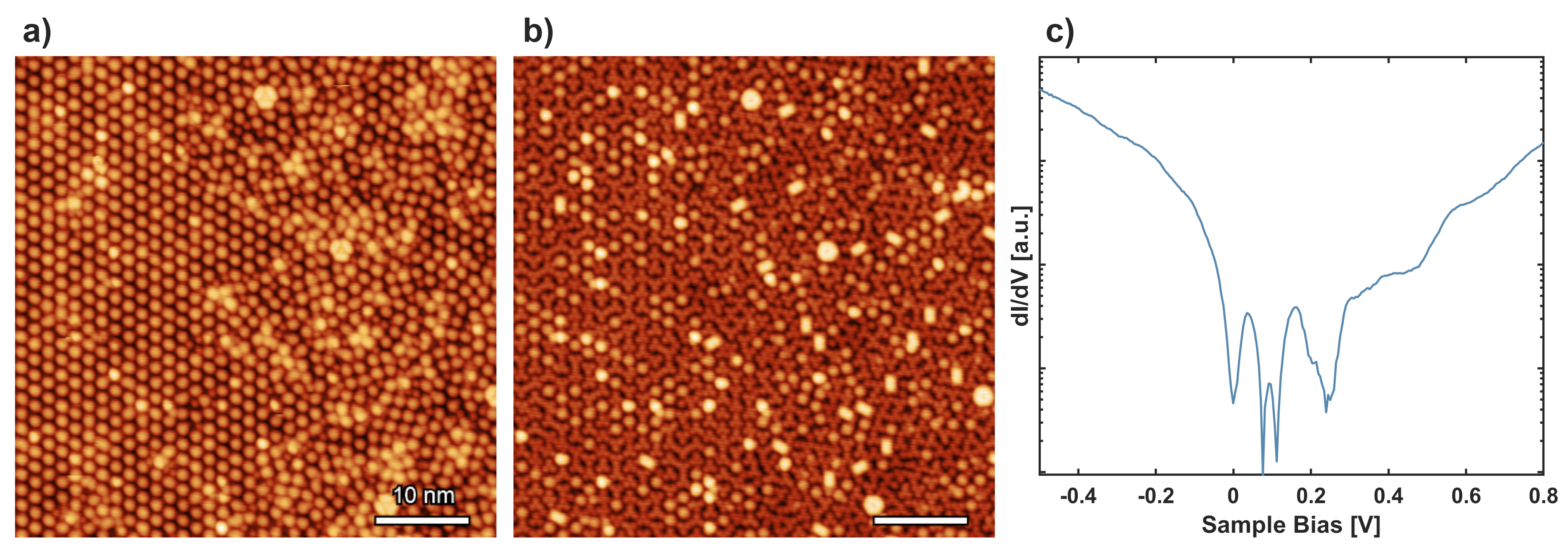}
\caption{\label{fig:4}(a) Empty state STM image of the InSb(111)B (3x1) surface reconstruction (V = 1.0 V, I = 74 pA). Hexamers are visible on the surface. A region of ordered Hexamers forming a (3x3) reconstruction and a region of disordered hexamers are present. (b) Filled state STM image of the same region, individual Sb atoms of the hexamers are visible (V = -1.0 V, I = 70 pA). (c) STS spectra of the (3x1) surface in an disordered region.}
\end{figure*}

The InSb(111)B (3x1) surface reconstruction is an In rich surface which has been observed in reflection high energy electron diffraction during thin film growth of InSb(111)B \cite{Noreika1981}. An empty and filled state image of the same region on (3x1) surface is shown is Fig. \ref{fig:4}(a) and (b), respectively. From the empty state image two distinct regions are present: a region of the previously observered (3x3) surface reconstruction and a region of disordered hexamers. It is apparent that this (3x1) surface is a mixed reconstruction surface composed of ordered and disordered hexamers. Sb atoms of the hexamers are visible in the filled state image of the same region. In this image the three Sb atoms of the (3x3) hexamer are visible and ordered. In the disordered region, the Sb atoms of the hexamer are rotated and often missing Sb atoms. To conclude, the (3x1) surface is a disordered surface comprised of hexamers which are rotated and often contain missing Sb atoms. This (3x1) surface is also a metastable surface, which will eventually convert to the (3x3) surface after annealing. The observed (3x1) surface in this study is very different than a prior study of the (111)B (3x1) surface, which reveals an ordered (3x1) surface reconstruction\cite{Bjorkqvist1996}. This difference is attributed to a difference in surface preparation. The ordered (3x1) surface was prepared by sputter annealing, whereas this surface was prepared by molecular beam epitaxy. The STS spectra of the disordered region is shown in Fig. \ref{fig:4}(c). In this spectra, multiple subgap states are observed, with the peak energies located at $47 \pm 25$, $112 \pm 25$, and $139 \pm 25$ mV. These surface states likely originate from the dangling bonds created from the missing atoms in the hexamers. The valence band edge and conduction band edges are located at $15 \pm 25$ and $246 \pm 25$ mV above the Fermi-level. For the (3x1) surface, the surface Fermi-level is pinned at the valence band edge.

\begin{figure}[]
\includegraphics[width=0.5\textwidth]{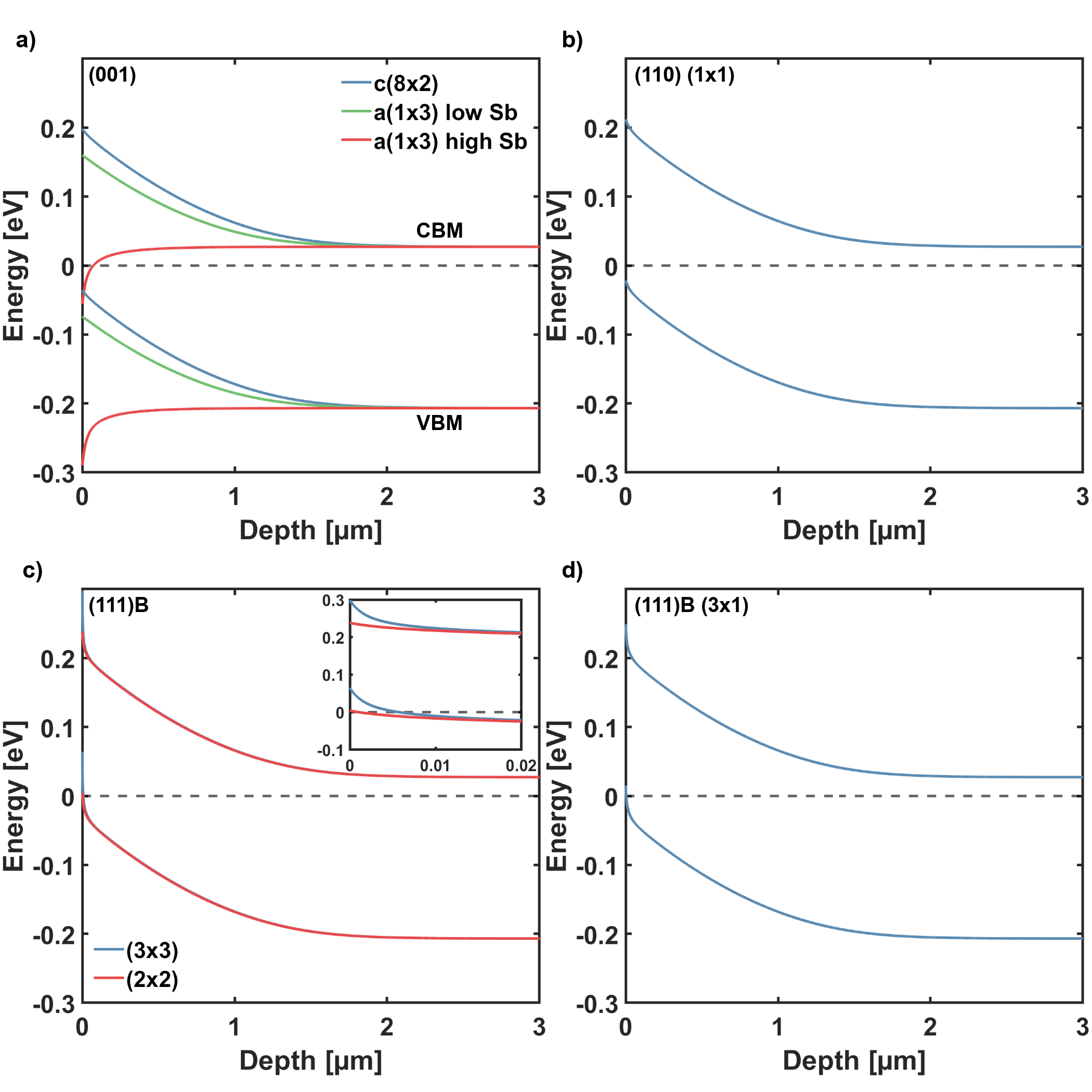}
\caption{\label{fig:5} Calculated surface band bending profiles of the various InSb surfaces. (a) The profiles for the (001) surfaces. (b) The profile of the (110) (1x1) surface. (c) The profile of the (111)B (2x2) and (3x3) surface, inset is the surface band bending region within 20 nm of the surface. (d) The band bending profile of the (111)B (3x1) surface.}
\end{figure}

The valence band edge energies ($E_v$), conduction band energies ($E_c$), and band gaps ($E_g$) of all of the measured surfaces are sumarized in Table \ref{tab:table1}. The band bending profiles of the various surfaces is shown in Fig. \ref{fig:5}, and the band bending profiles were computed using the method described by \textit{Tan et al.}\cite{Tan1990}. The Fermi-level of the In rich and the equal In:Sb surfaces is pinned near the valence band maximum and upwards band bending occurs at the surface. This observed Fermi-level pinning is consistent with previous reports of a charge neutrality level near the valence band and acceptor like surface states \cite{Etchells1976,Forster1987,Akkal2000,King2008}. Upon deposition of Sb on the (001) and (111)B surface Fermi-level shifts upwards towards the conduction band. In the (001) a(1x3) high Sb surface, downward band bending occurs, and the surface is in electron accumulation. We propose the origin of the upwards shift of the surface Fermi-level upon Sb deposition is electron doping of the surface from Sb clusters, which are present on the Sb rich surfaces. Sb clusters can be semimetallic
\cite{Patrin1992}, and can potentially serve as an electron donor on the surface. In comparison to the (001) surface, excess Sb on the (111)B surface does not appear to shift the surface Fermi-level as significantly. The smaller measured Fermi-level shift it potentially due to a higher surface state density on the (111)B surface.

\begin{table}[]
\caption{\label{tab:table1} Valence band edge energy ($E_v$), conduction band edge energy ($E_c$), and band gap energy ($E_g$) of the various InSb surfaces. All band edge energies are with respect to the Fermi-level. Band gap energy is defined as: $E_g = E_c - E_v$.}
\begin{ruledtabular}
    \begin{tabular}{cccc}
         Surface & $E_v$ (meV) & $E_c$ (meV) & $E_g$ (meV)   \\
         \hline
         (001) c(8x2)  & $-36 \pm 25$ & $213 \pm 25$ & $249 \pm 35$   \\ 
         (001) a(1x3)\footnotemark[1]  & $-74 \pm 25$ & $197 \pm 25$ & $271 \pm 35$   \\ 
         (001) a(1x3)\footnotemark[2] & $-290 \pm 25$ & $-63 \pm 25$ & $227 \pm 35$   \\
         (110) (1x1)  & $-22 \pm 25$ & $249 \pm 25$ & $271 \pm 35$   \\ 
         (111)B (2x2)  & $4 \pm 25$ & $275 \pm 48$ & $271 \pm 54$   \\ 
         (111)B (3x3)  & $64 \pm 25$ & $285 \pm 25$ & $221 \pm 35$   \\ 
         (111)B (3x1) & $15 \pm 25$ & $256 \pm 25$ & $231 \pm 35$   \\ 
         
    \end{tabular}
    \end{ruledtabular}
    \footnotetext[1]{Low Sb} \footnotetext[2]{High Sb}
\end{table}

\section{Conclusions}

The electronic structure of clean InSb (001), (110), and (111)B surfaces was studied with STM/S. We find that the Sb shifts the surface Fermi-level towards the conduction band on the (001) surface, and Sb exposure to the surface can shift the surface from depletion to electron accumulation. In contrast to the (001) surface, excess Sb on the (111)B surface does not as effectively shift the surface Fermi-level, potentially due to a higher acceptor surface state density. The surface Fermi-level of the In rich and equal In:Sb (001), (110), and (111)B surfaces remains close to the valence band maximum, which is consistent with a charge neutrality level near the valence band maximum. We determined that the (111)B (3x1) surface is a metastable, disordered (3x3) surface with many vacancies. Multiple surface states were present within the band gap of the (3x1) surface. These surface states are attributed to the disorder and the vacancies present in the surface. These findings yield further insights into the Fermi-level pinning of various InSb surfaces and demonstration of Fermi-level tuning of (001) and (111)B surfaces devices via Sb terminated surfaces, thereby enabling further tunability of InSb based devices.


%
%

%


\begin{acknowledgments}
This work was supported by Microsoft Station Q and University of California Multiple Campus Award No. 00023195. We would like to acknowledge valuable and fruitful discussion with G. W. Winkler, M. Thomas, R. M. Lutchyn, and S. Nadj-Perje.
\end{acknowledgments}

\section*{Author Declarations}

\subsection*{Conflict of Interest}
The authors have no conflicts to disclose.

\section*{Data Availability}
The data that support the findings of this study are available
from the corresponding author upon reasonable request.


\section*{References}
\bibliography{InSbPaper}

\end{document}